\documentclass[a4paper,twocolumn]{esapub2005} 
\usepackage{times}
\usepackage{epsfig}
\usepackage{amssymb,natbib}

\input cyracc.def
\font\tencyr=wncyr10
\def\cyr{\tencyr\cyracc}

\def\cm{\mbox{ cm}}
\def\cmmoinsdeux{\mbox{ cm}^{-2}}
\def\cmmoinstrois{\mbox{ cm}^{-3}}

\def\microns{\mbox{ } \mu \mbox{m}}

\def\mags{\mbox{ magnitudes}}

\def\ergpars{\mbox{ erg\,s}^{-1}}

\def\adeg{^{\circ}}

\def\amin{^\prime}

\def\asec{^{\prime \prime}}

\def\nh{N_{\rm H}}

\def\ltsima{\; \buildrel < \over \sim \;}
\def\simlt{\lower.5ex\hbox{\ltsima}}            
\def\gtsima{\; \buildrel > \over \sim \;}
\def\simgt{\lower.5ex\hbox{\gtsima}}            

\title{Optical to Mid-Infrared observations revealing the 
most obscured high-energy sources of the Galaxy.}
\author{Sylvain Chaty}
\affil{
AIM - Astrophysique Interactions Multi-\'echelles 
(Unit\'e Mixte de Recherche CEA/CNRS/Universit\'e 
Paris 7 Denis Diderot n$\adeg$7158),
CEA Saclay, DSM/DAPNIA/Service d'Astrophysique, B\^at. 709,
L'Orme des Merisiers, FR-91 191 Gif-sur-Yvette Cedex, France, 
{\em chaty@cea.fr}}
\author{Farid Rahoui}
\affil{European Southern Observatory, Alonso de Cordova 3107, Vitacura,
Santiago de Chile \&
AIM - Astrophysique Interactions Multi-\'echelles 
(UMR CEA/CNRS/Universit\'e Paris 7 Denis Diderot n$\adeg$7158), France, 
{\em frahoui@eso.org}}

\begin{document}

\keywords{{\it INTEGRAL}; IGR J16195-4945; IGR J16318-4848; IGR J17544-2619; near-infrared; mid-infrared}

\maketitle

\begin{abstract}
A new type of sources has been discovered by {\it INTEGRAL}.  These sources
are in the course of being unveiled by means of multi-wavelength
optical, near- and mid-infrared observations. Among the high-energy
binary sources, two distinct classes are appearing. The first class is
constituted of intrinsically 
obscured high-energy sources, of which IGR J16318-4848
seems to be the archetype. The second class is populated by the
so-called supergiant fast X-ray transients, with IGR J17544-2619
being the archetype. We report here on multi-wavelength
observations of sources from these two classes, focusing on optical
to mid-infrared observations. We show that in the case
of the obscured sources IGR J16318-4848 and IGR J16195-4945,
our observations suggest the presence of absorbing 
material (dust and/or cold gas) enshrouding the whole binary system.
We then discuss the nature of these two different types of sources.
\end{abstract}

\section{Introduction}

{\it INTEGRAL} has performed a detailed survey of the galactic plane
and the ISGRI detector on the IBIS imager has discovered many new
sources, most of all 
reported in \cite{bird:2006}\footnote{Updated informations about
these sources are reported in  http://isdc.unige.ch/~rodrigue/html/igrsources.html}.  Many of these new sources are
concentrated in a direction tangent to the Norma arm of the Galaxy
(see e.g. \citeauthor{chaty:2005a} \citeyear{chaty:2005a} and
\citeauthor{tomsick:2004} \citeyear{tomsick:2004}), a region of our Galaxy
which is rich in star forming regions.  The most important result of
the {\it INTEGRAL} observatory to date is probably the discovery of
many new high energy sources exhibiting common characteristics which previously
had rarely been seen. Most of them are high mass X-ray binaries (HMXBs)
hosting a neutron star orbiting around an O/B companion, in some cases
a supergiant star. They then divide into two classes. Some of the new
sources are very obscured, exhibiting a huge intrinsic and local
extinction.  The archetype, and certainly the best example, is the
extremely absorbed source IGR J16318-4848
\citep{filliatre:2004}.  The other sources are HMXBs hosting a
supergiant star, exhibiting fast and transient outbursts: an
unusual characteristic among HMXBs.  They are therefore called
Supergiant Fast X-ray Transients (SFXTs, \citeauthor{negueruela:2006},
\citeyear{negueruela:2006}).

Since high-energy observations are not enough in order to reveal the
nature of the newly discovered sources, one needs to perform
multi-wavelength observations.  Indeed, the difficulty is that even if
{\it INTEGRAL} can provide a position for these sources which is
already very accurate for this energy range ($\sim 2\amin$), the
localisation is not accurate enough in order to pinpoint the source at
other wavelengths. So the first stage is to observe in the low energy
part of the high energy domain, for example with {\it XMM-Newton} or
{\it Chandra}. These satellites can give arcsecond position
accuracy. At this stage, the hunt for the optical counterpart of the
source is open.  However, once again, there is a difficulty, due to
the high level of absorption in this region of the Galaxy, close to
the galactic plane.  One has then to observe in the near-infrared
(NIR) domain in order to begin to reveal these sources at these
wavelengths.  Furthermore, since there is a strong absorption, there
must obviously be some absorbing matter...  It is only by observing at
mid-infrared (MIR) wavelengths that one can characterise the nature of
this absorbing matter, and determine if it is made of cold gas, or
dust, or anything else...

We first report here on multi-wavelength observations of the two archetypes
described above, and give results on MIR observations of
 newly discovered {\it INTEGRAL} sources belonging to both classes,
in Section \ref{observations}. We then
 discuss these results and conclude in Section \ref{discussion}.

\section{Observations and results} \label{observations}

In order to study the newly discovered {\it INTEGRAL} sources, it is
necessary to perform multi-wavelength observations in the optical, NIR
and MIR domains.
The multiwavelength  observations that we describe here were
performed at the European Southern Observatory (ESO), in 3 domains:

$\bullet$ optical observations ($400-800 \microns$) with the EMMI instrument, 
on the 3.5m New Technology Telescope (NTT) at La
Silla, \\
$\bullet$ NIR observations ($1-2.5 \microns$) with the SOFI
instrument, on the NTT, \\
$\bullet$ and MIR observations ($5-20 \microns$)
 with the VISIR instrument on Melipal, the
8m third Unit Telescope (UT3) of the Very Large Telescope (VLT) at
Paranal.

These observations have been done using two different modes: 
Target of Opportunity (ToO) and Visitor modes.
They include photometry and spectroscopy on 15 {\it INTEGRAL} sources
in order to identify their counterparts, the nature of the
companion star, derive the distance, and finally characterise 
the presence, the temperature, and the composition of their
circumstellar medium.

     \subsection{IGR J16318-4848: the archetype of the obscured high-energy
sources}

We will first remind the main characteristics of this source in the
high energy domain (mainly reported in \citeauthor{matt:2003} 
\citeyear{matt:2003} and \citeauthor{walter:2003} \citeyear{walter:2003}), 
before describing the optical/NIR observations of
this source.  IGR J16318-4848 was the first source to be discovered by
the ISGRI detector on the IBIS imager onboard {\it INTEGRAL}, on 29
January 2003 at the galactic coordinates $(l,b) \sim (336\adeg,
0.5\adeg$), with an uncertainty radius of localisation of
$2\amin$ \citep{courvoisier:2003}.  
ToO observations were then triggered with {\it XMM-Newton},
which allowed a more accurate localisation at $4\asec$.  {\it
XMM-Newton} observations showed that the source was exhibiting a
strong absorption of $\nh \sim 2 \times 10^{24} \cmmoinsdeux$, a
temperature of kT~$= 9$~keV, and a photon index $\sim 2$.  In the high
energy spectrum a strong Fe absorption edge was visible, altogether
with Fe K$\alpha$, K$\beta$ and Ni K$\alpha$ fluorescence emission
lines.  The 15-40 keV flux was 50-100 mCrab, the luminosity (assuming
that the source is located at 5 kpc) was $L_{5kpc} = 1-20 \times 10^{36}
\ergpars$.  The flux was highly variable (by a factor of 20), and no
oscillation was detected.  There were usually 10 hours between flares,
and 2 to 3 days of inactivity were also observed.  The lines and
continuum were varying on a 1000 s timescale: this allowed to derive
the size of the emitting region to be smaller than $3 \times 10^{13} \cm$.
These X-ray properties, signature of wind accretion, 
were reminiscent of other peculiar sources, such as 
CI Cam and GX 301-2.

The accurate localisation allowed us to look for the counterpart at 
other wavelengths; the results that we will now report come
from \cite{filliatre:2004}, and we refer to this paper for more 
details. 
ToO photometrical and spectroscopic observations in optical and NIR
were triggered just after the discovery of the source, but the observations
could not be performed before 23-25 February 2003.
\cite{walter:2003} had reported the discovery of the
optical and NIR counterpart, however, after an improved astrometry
based on these new optical/NIR observations, \cite{filliatre:2004}
showed that they had misidentified the optical counterpart. Two optical
sources were present inside the {\it XMM-Newton} EPIC $4\asec$
uncertainty circle, but comparison with the USNO B1.0 plate R band showed
that only one of the two sources varied.  This independent and
improved astrometry therefore allowed
\cite{filliatre:2004} to discover the real optical counterpart, 
and to confirm the NIR counterpart proposed by \cite{walter:2003}.
The optical counterpart of the source is not seen in the B and V
filters (B~$> 25.4 \pm 1.0$; V $> 21.1 \pm 0.1$), and appears in the
R, I and Z filters (R $= 17.72 \pm 0.12$; I $= 16.05 \pm 0.54$).  The
first striking fact in the optical/NIR observations was the extreme
brightness of this source in the NIR: the magnitudes of the NIR
counterpart were J $= 10.33\pm 0.14$; H $=8.33\pm 0.10$ and Ks
$=7.20\pm0.05$. This source is too bright to perform photometrical
observations with a 4m class telescope, even with an integration time
of 1s! This shows the need to maintain small telescopes for such
observations.

The second striking fact in the optical/NIR domain is the absorption
in the optical of this source.  By looking at the magnitude versus the
optical/NIR (B, V, R, I, J, H, Ks) wavelengths of IGR J16318-4848 and
neighbour objects in the field of view, it is obvious that the high
energy source exhibits an unusually strong intrinsic absorption in the
optical of $A_v = 17.4 \mags$, much stronger than the absorption along
the line of sight as exhibited by the neighbour objects (absorption of
$A_v = 11.4 \mags$), but still 100 times lower than the absorption in
X-rays!
This led \cite{filliatre:2004} to suggest 
that the material absorbing in the X-rays 
must be concentrated around the compact object, while there is some
material absorbing in the optical/NIR which is concentrated around
the whole system.

\begin{figure}
  \centerline{\includegraphics[width=9.cm]{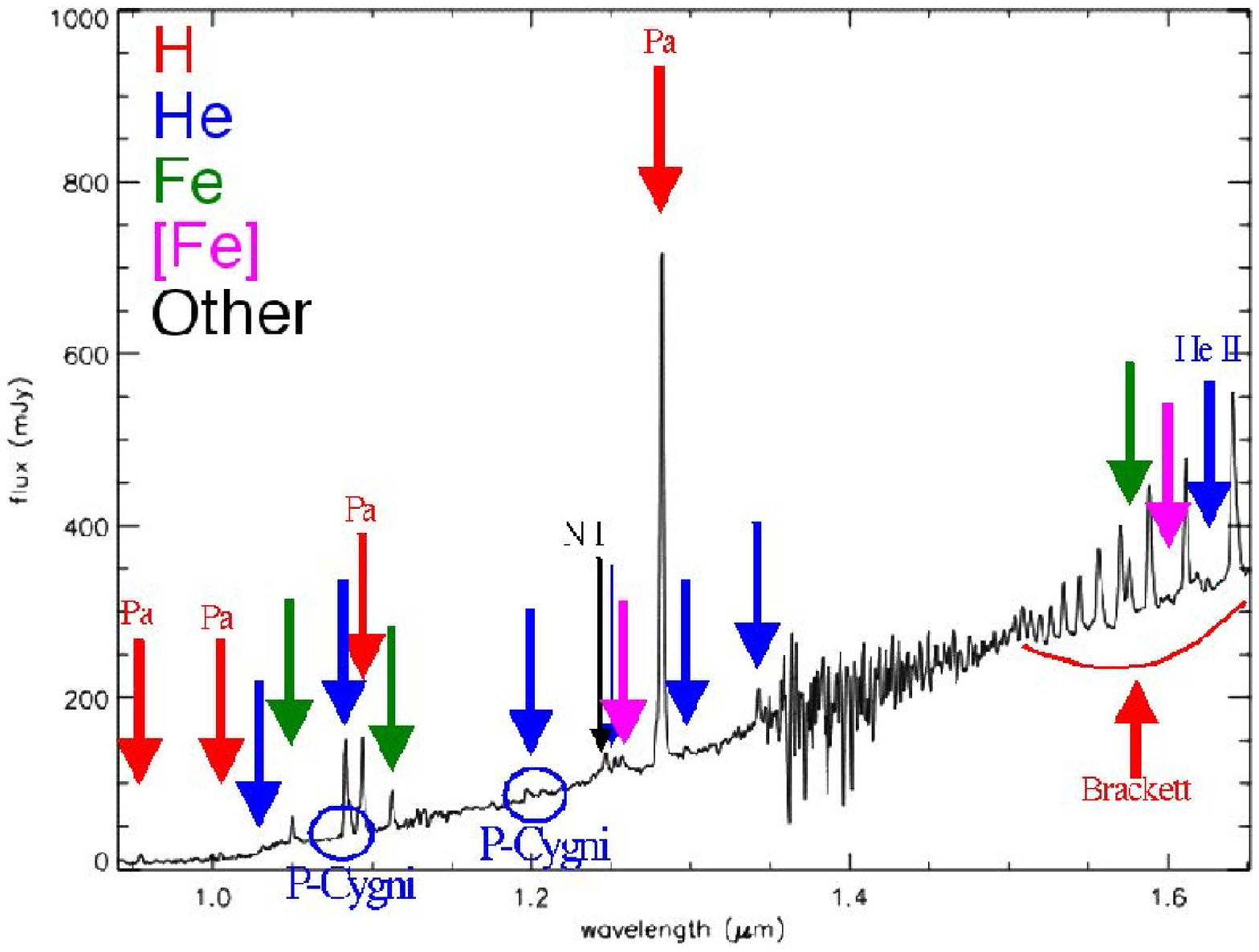}}
  \caption{NIR spectrum (0.95-1.65 $\microns$) of IGR J16318-4848,
taken at ESO/NTT \citep{filliatre:2004}.}
  \label{figure:spec1}
\end{figure}

\begin{figure}
  \centerline{\includegraphics[width=9.cm]{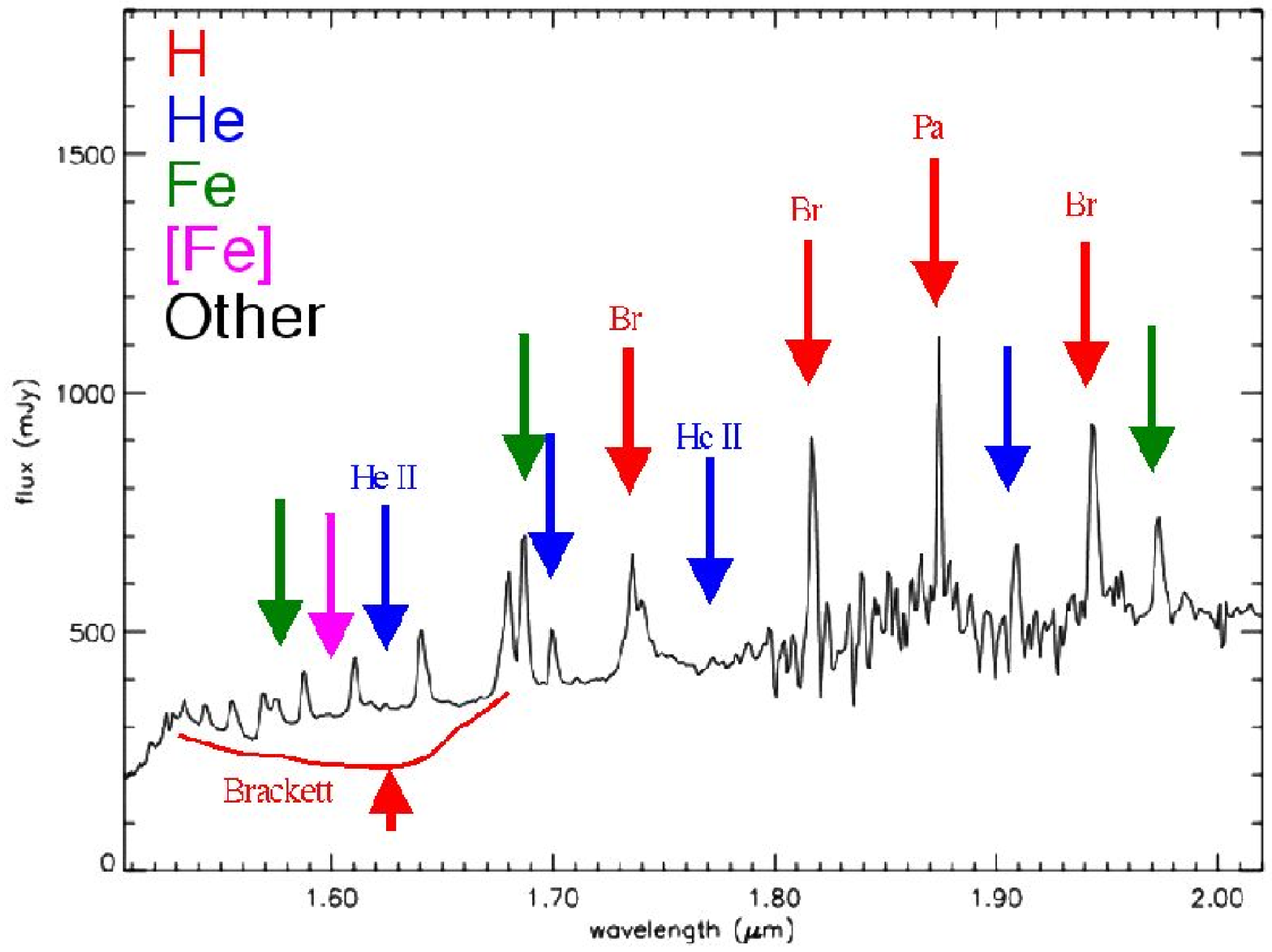}}
  \caption{NIR spectrum (1.5-2.05 $\microns$) of IGR J16318-4848,
taken at ESO/NTT \citep{filliatre:2004}.}
  \label{figure:spec2}
\end{figure}

\begin{figure}
  \centerline{\includegraphics[width=9.cm]{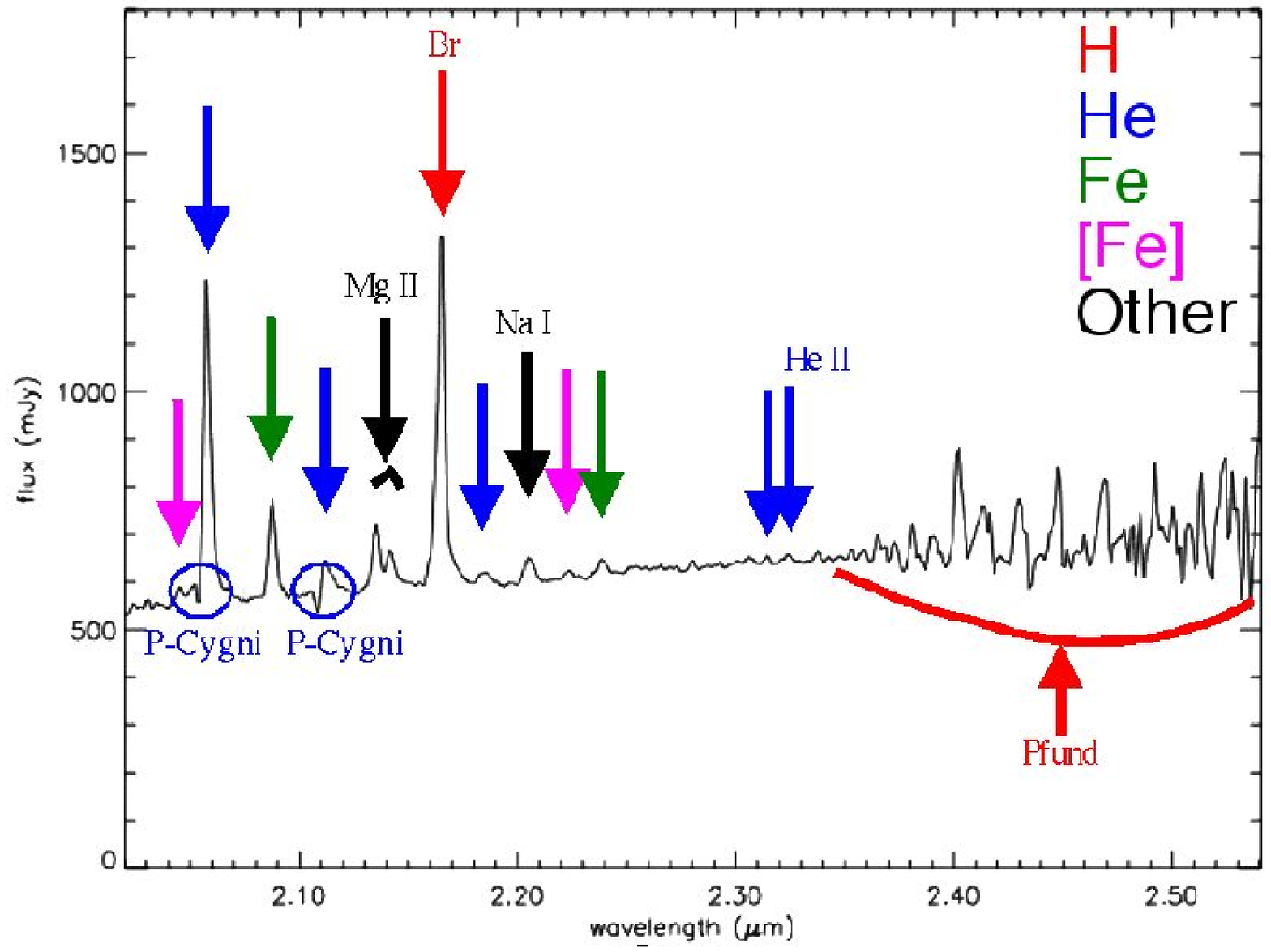}}
  \caption{NIR spectrum (2.0-2.55 $\microns$) of IGR J16318-4848,
taken at ESO/NTT \citep{filliatre:2004}.}
  \label{figure:spec3}
\end{figure}

The NIR spectroscopy in the $0.95-2.5 \microns$ 
domain, shown in Figures \ref{figure:spec1}, \ref{figure:spec2}, and 
\ref{figure:spec3}, 
revealed the third striking fact of this optical/NIR domain:
the high energy source exhibits an unusual spectrum in NIR, very rich
in many strong emission lines.  The different lines allow us to
characterise the medium around this object. We remind here the main
characteristics:

$\bullet$ the strong H (Brackett, Paschen, Pfund) 
and HeI (P-Cygni profiles) lines emanate from a dense and ionised wind,\\
$\bullet$ the He II lines come from a highly excited region 
in the vicinity of the compact object,\\
$\bullet$ the forbidden [FeII] lines indicate the presence
of shock heated material,\\
$\bullet$ the allowed FeII lines imply a medium with densities greater 
than $10^5-10^6 \cmmoinstrois$,\\
$\bullet$ the NaI lines arise from cool and dense regions, shielded from both 
stellar and compact object radiation.


From these characteristics it is therefore clear that lines originate
from different media (exhibiting various densities and temperatures),
suggesting the presence in this high energy source of a highly complex
and stratified circumstellar environment, and also the presence of an
envelope and a wind.  Only luminous post main sequence stars show such
extreme environments, and then the companion star is most likely 
a sgB[e] star (or even an unclassified uncl/sgB[e] 
star because of its high luminosity).  
The system is therefore a high-mass X-ray binary
system.  As it was the case in the X-rays, the NIR characteristics are
also reminiscent of the other peculiar high energy source CI Cam.

\cite{filliatre:2004} then built the multi-wavelength 
Spectral Energy Distribution (SED) of this source, from radio to
X-rays, including optical and NIR domains.  They fitted a black-body
representing the companion star to these observations, and derived the
following parameters: A$_v = 17.5 \mags$, L~$\sim 10^6 D_{6 kpc}^2
\times L_{\odot}$, T~$= 20 250$~K, M~$= 30 M_{\odot}$ and $r/D = 5\times
10^{-10}$, where A$_v$ is the absorption in the V band and L, T, M, r and
D are the companion star luminosity, temperature, mass, radius and distance
respectively.  These parameters imply a high luminosity, high temperature
and massive star, therefore likely a supergiant, located at 6 kpc.
The photometry, spectroscopy and fit of the SED therefore give results
which are consistent between each other.  Furthermore, by locating
these parameters on a Hertzprung-Russel (or temperature--luminosity)
diagram, one can see that this companion star is located at the edge
of the blue supergiant domain, indicating that we are facing an extreme
object even among those already extreme blue supergiant stars!

The SED can also allow us to try to derive the nature of the compact object.
Indeed, a correlation in the black hole systems 
associated with low/hard X-ray emission has been found
between X-ray and radio flux densities \citep{gallo:2003}.
If the compact object of IGR J16318-4848 were a black hole,
its 50-100 mCrab low/hard X-ray flux would lead to a 10 mJy radio flux.
However, radio ATCA observation on 9 February 2003 did not detect any source
 up to 0.1 mJy, suggesting that the compact object is a neutron star.
But we point out that we have to be cautious, since
this correlation might not be so universal,
see for instance \cite{cadolle-bel:2006}.

\begin{figure}[t]
\centerline{\psfig{file=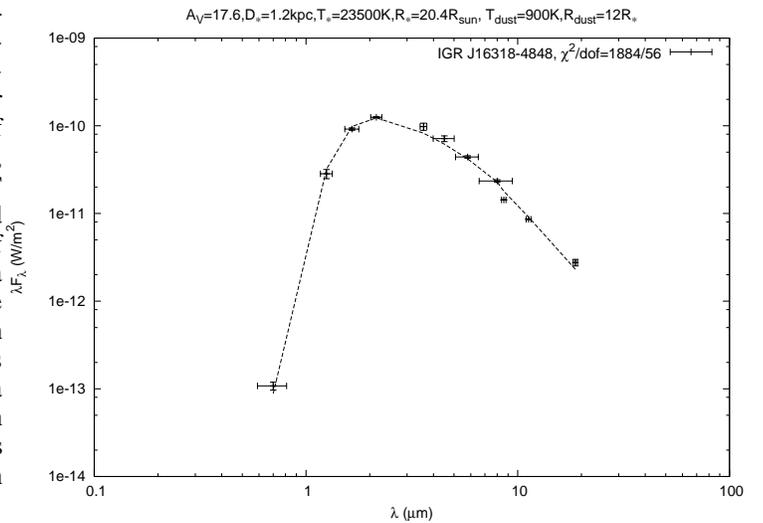,angle=-90.,width=10.cm}}
\caption[]{\label{figure:igrj16318} Optical to MIR SED of IGR J16318-4848,
including data from ESO/NTT, VISIR on VLT/UT3 and {\it Spitzer} (GLIMPSE survey). 
The observations fitted with both a model of a companion star 
(taking usual parameters of a sgB[e]) and simple spherical dust component
allowed us to derive the following parameters:
i) for the companion star: temperature
T $=23500 K$ and radius R$_{star} = 20.4 R_{\odot} = 15
\times 10^6$ km;
ii) for the dust component: temperature T $=900K$, radius
R $= 12 R_{star} = 171 \times 10^6$ km.
The absorption we derived was A$_v = 17.6 \mags$, and the distance
 D $=1.2$ kpc. The high $\chi^2$/dof value of the fit (=1884/56)
is mainly due to the very small error bars in the MIR domain.  
The main result given by this SED is that 
in the case of IGR J16318-4848 there is a need for an
extra (e.g. dust) component in order to fit its SED. }
\end{figure}

Now, the question which remains open is: what is the cause of this
unusual absorption in the optical/NIR domain? In order to answer to this
question, we recently obtained MIR observations with VISIR on VLT/UT3.
We were therefore able to fit IGR J16318-4848 SED from optical to MIR
wavelengths, including data from ESO/NTT, {\it Spitzer} (GLIMPSE survey) and
VLT/VISIR (see Figure \ref{figure:igrj16318}).  We fitted the
observations with a model of a companion star (taking usual parameters
of a sgB[e]) and simple spherical dust component. More details on this fitting
procedure are given in Section \ref{MIR}.  
We found for the parameters of the companion star a temperature of
T\,$=23500$\,K, radius R$_{star} = 20.4 R_{\odot} = 15
\times 10^6$\,km,
and a dust component with the following parameters: T $=900$\,K, radius
R\,$= 12 R_{star} = 171 \times 10^6$\,km and A$_v = 17.6 \mags$.  The
derived distance was of D~$=1.2$~kpc.  The $\chi^2$/dof of the fit was
of 1884/56, a high value mainly due to the very small
error bars in the MIR domain (more details about this work will
be given in Rahoui and Chaty, in prep.).  What is important in this result is
that in the case of IGR J16318-4848 there is a need for
an extra (e.g. dust) component. The extension of this dust component
seems to suggest that it enshrouds the whole binary system,
perhaps as would do a cocoon of dust.

Let us now summarise briefly the nature of IGR J16318-4848.  We are
facing an HMXB system, located at a distance between 1 to 6 kpc,
hosting a neutron star (probably) and an early-type supergiant B[e]
star.  It is therefore the second HMXB with a sgB[e] star, after CI Cam.  The
most striking facts are that i) the compact object seems to be
surrounded by absorbing material and ii) the whole system itself
seems to be surrounded by a dense and absorbing dusty circumstellar
material envelope or cocoon, and by both cold and hot stellar wind components.

     \subsection{IGR J17544-2619: the archetype of the Supergiant Fast
X-ray Transients}

The Supergiant Fast X-ray Transients (SFXTs) is a class of sources
identified among the recently discovered {\it INTEGRAL} sources.  As
their name indicates it, this class is constituted of high-energy
transient sources, whose common characteristics are: they exhibit
rapid outbursts, lasting only hours, a faint quiescent emission, their
high energy spectra require a BH or NS accretor, and they host O/B
supergiant companion stars.  Among these sources, IGR J17544-2619
seems to be the archetype, since it exhibits all characteristics that
are common to sources belonging to this SFXT class.  We will now review
high energy properties of IGR J17544-2619 before reporting optical/NIR
observations.

IGR J17544-2619 is a bright recurrent transient X-ray source which has
been discovered by {\it INTEGRAL} on 17 September 2003, at $3\adeg$
from the Galactic centre \citep{sunyaev:2003}.  {\it XMM-Newton} had
observed the field of this source, and EPIC had detected an X-ray
counterpart with mean 0.5-10 keV unabsorbed variable luminosity of
$1.1-5.7 \times 10^{35} \ergpars$ for an assumed distance of 8 kpc (see
\citeauthor{gonzalez-riestra:2004} \citeyear{gonzalez-riestra:2004}).  
The EPIC spectra can be represented by a power-law model with variable
photon indices of $1.42-2.25 \pm 0.15$. The 0.5-10 keV spectrum
hardens with increasing intensity.  This source has therefore a very
hard X-ray spectrum, and exhibits a faint intrinsic absorption
($10^{22} \cmmoinsdeux$).  Its bursts last for hours, in-between
bursts it exhibits long quiescence periods, and there is a long
outburst period of 165 days \citep{negueruela:2006}.  Furthermore,
there is no radio emission up to an upper limit of 7.35 mJy at 0.61 GHz
\citep{pandey:2006}.  The nature of the compact object is probably a
neutron star, as suggested by \cite{in'tzand:2005}.  Some of the
high-energy properties of this source were similar to other {\it
INTEGRAL} sources (such as IGR J16318-4848, IGR J16320-4851, IGR
J16358-4726, see e.g. \citeauthor{chaty:2005a}
\citeyear{chaty:2005a}).
The question which
rapidly arose concerning this source was then: is it another highly
absorbed galactic X-ray binary?  But the rapid outbursts of this source,
quite unusual among the HMXBs, could even 
suggest that it was belonging to a new
kind of X-ray binaries. It was therefore important to establish its
nature, and once again, the optical and infrared observations will
play here a crucial role in revealing this source.

\cite{pellizza:2006} managed to get optical/NIR ToO observations only one day
after the discovery of this source. We report here their main results,
and refer to \cite{pellizza:2006} for more details on the study of the
optical/NIR counterpart of this source.
Inside the {\it INTEGRAL} $2\amin$ uncertainty radius circle of IGR
J17544-2619, there is a ROSAT source (1RXS J175428, with $23\asec$
uncertainty radius) which is in fact not connected with IGR
J17544-2619.  {\it XMM-Newton} observations have allowed to accurately
localise the source with a $4\asec$ uncertainty radius circle. But
even in such a small circle, there were 5 optical candidate
counterparts inside: a bright candidate (called C1 in
\citeauthor{pellizza:2006} \citeyear{pellizza:2006}) 
identified in USNO and 2MASS, three very faint candidates (C2, C3 and C5), 
probably foreground dwarf stars, and finally an extended object (C4), likely a
high-z galaxy.
\cite{pellizza:2006} have shown that the C1 candidate had to be the 
counterpart of IGR J17544-2619, 
and confirmed it by an astrometry including {\it Chandra}
observations. Indeed, these observations allowed to localise this
source even more accurately, with a $0.4\asec$ uncertainty error
circle.

\begin{figure}
\resizebox{\hsize}{!}{\includegraphics{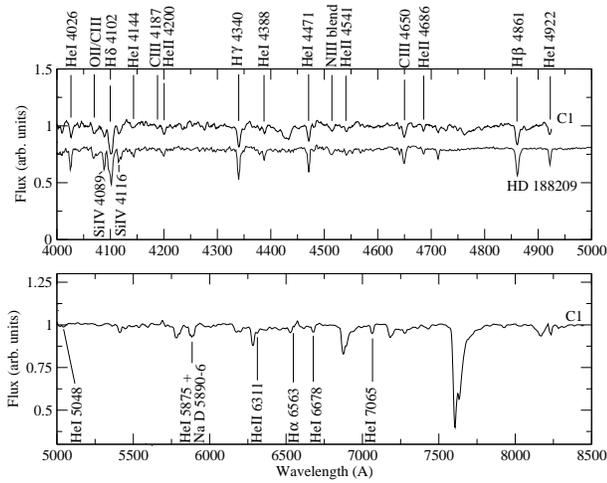}}
\caption{Optical spectrum of IGR J17544-2619 showing the identified
lines, among which \cite{pellizza:2006} found strong HI, HeI and HeII
lines typical of an O-type star. {\bf Upper panel:} Blue spectrum of 
IGR J17544-2619 (upper
curve) and the standard O9Ib star HD 188209. The high degree of
similarity between them supports their 
classification of IGR J17544-2619 as an O9Ib star. {\bf
Lower panel:} Spectrum of IGR J17544-2619 between 5000~{\AA} and 8500~{\AA}.}
\label{figure:spectrum}
\end{figure}

Spectroscopy was thereafter performed on the C1 candidate, and the
spectrum is shown in Figure \ref{figure:spectrum}, characteristic of a
blue supergiant of spectral type O9Ib, with a mass of $25-28
M_{\odot}$ and temperature of T $\sim 31000$ K: the system is
therefore an HMXB.  The spectrum also exhibits $H\alpha$ P-Cygni
profiles, suggesting the presence of a stellar wind at a velocity of $265
\pm 20$~km/s. If this is confirmed, it 
is unusually mild for O stars (this value is for
instance smaller than the wind velocity of 400 km/s in IGR J16318-4848
reported in \citeauthor{filliatre:2004} \citeyear{filliatre:2004}).
The derived distance was of 3-4 kpc.

\begin{figure}
\centerline{\psfig{file=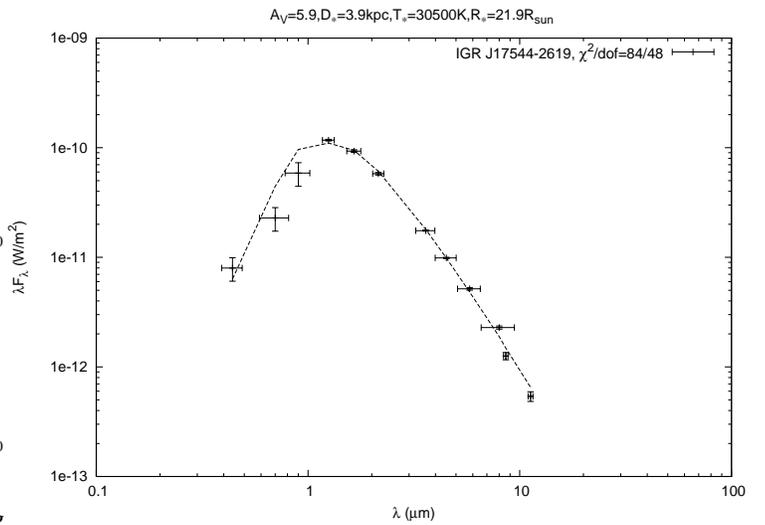,angle=-90.,width=10.cm}}
\caption[]{\label{figure:igrj17544} Optical to MIR SED of IGR J17544-2619,
including data from ESO/NTT, VISIR on VLT/UT3 and {\it Spitzer} (GLIMPSE survey). 
The observations were fitted with a model of a companion star 
(taking usual parameters of an O/B star) 
which allowed us to derive the following parameters for the companion star: 
temperature T~$=30500$~K and radius R$_{star} = 21.9 R_{\odot} = 15
\times 10^6$~km;
The absorption we derived was A$_v = 5.9 \mags$, and the distance
 D~$=3.9$~kpc. The $\chi^2$/dof value of the fit is 84/48.
The main result given by this SED is that 
in the case of IGR J17544-2619 there is no need for any
extra (e.g. dust) component in order to fit its SED, and only
a stellar component is necessary.}
\end{figure}

In order to better characterise the nature of the emission of this
source, and to answer to the question of the presence of dust around
this type of sources, we recently obtained MIR observations with VISIR
on VLT/UT3.  We were therefore able to fit IGR J17544-2619 SED from
optical to MIR wavelengths, including data from ESO/NTT, {\it Spitzer}
(GLIMPSE survey) and VISIR on VLT/UT3 (see Figure
\ref{figure:igrj17544}).  We fitted the observations with a model of a
companion star (taking usual parameters of an O9Ib star) allowing us
to derive the following parameters for the companion star: temperature
T~$=30500$~K and radius R$_{star} = 21.9 R_{\odot} = 15 \times 10^6$~km.
The absorption we derived was A$_v = 5.9 \mags$, and the distance
D~$=3.9$~kpc. The $\chi^2$/dof value of the fit is 84/48
(more details about this work will
be given in Rahoui and Chaty, in prep.).
The main result given by this SED is that in the case of IGR
J17544-2619 there is no need for any extra (e.g. dust) component in order
to fit its SED, and only a stellar component is necessary.

Let us now summarise the nature of IGR J17544-2619. It is a HMXB at a
distance of 3-4 kpc, constituted by a supergiant of spectral type O9Ib
(Mass of $25-28 M_{\odot}$), with a mild stellar wind and a compact
object which is likely a neutron star, without any absorbing material.

     \subsection{MIR observations of {\it INTEGRAL} sources} \label{MIR}

Apart from the two sources described above, we also observed other
newly discovered {\it INTEGRAL} sources.  The optical and NIR results
will be presented in Chaty et al. (in prep.).  We report here the main
results about MIR observations (more details about this work will
be given in Rahoui and Chaty, in prep.).
The MIR observations were performed on 2005/2006 at Paranal UT3-VISIR, and
we present the results in Table \ref{tab:mirobs}. 
From this Table one can see that we detected in the MIR domain 9 {\it
INTEGRAL} sources out of 14.  
We fitted the optical to MIR
emission of these sources by an absorbed blackbody representing the stellar
emission. The free parameters of the fit were the absorption in the
V-band, the system distance and the companion star blackbody
temperature and radius.
The absorption at wavelengths $\lambda$ was computed using
$\frac{\textrm{A}_{\lambda}}{\textrm{A}_{V}}$ ratios given in 
\cite{rieke:1985} for optical bands and MIR wavelengths above $8 \microns$. 
For wavelengths from 1.25 to $8 \microns$, we used the
analytical expression given in \cite{indebetouw:2005}.  Fits were
optimised by minimising the ${\chi}^2$.
 
For most of the sources the SEDs were accurately fitted, showing
that the MIR emission has a stellar origin, and corresponds to the
Rayleigh-Jeans tail of the blackbody stellar spectrum.
However, in the case of IGR J16318-4848 (as described above) and
IGR J16195-4945, the fitted fluxes were too low compared to the
observed MIR fluxes.  We therefore had to add the
blackbody emission of a spherical dust cloud centred on the companion
star, in order to improve the fits in the MIR domain, considering
their high fluxes at these wavelengths. Only in these two cases,
dust cloud blackbody temperature and radius were also free parameters
of the fits.

\begin{figure}[t]
\centerline{\psfig{file=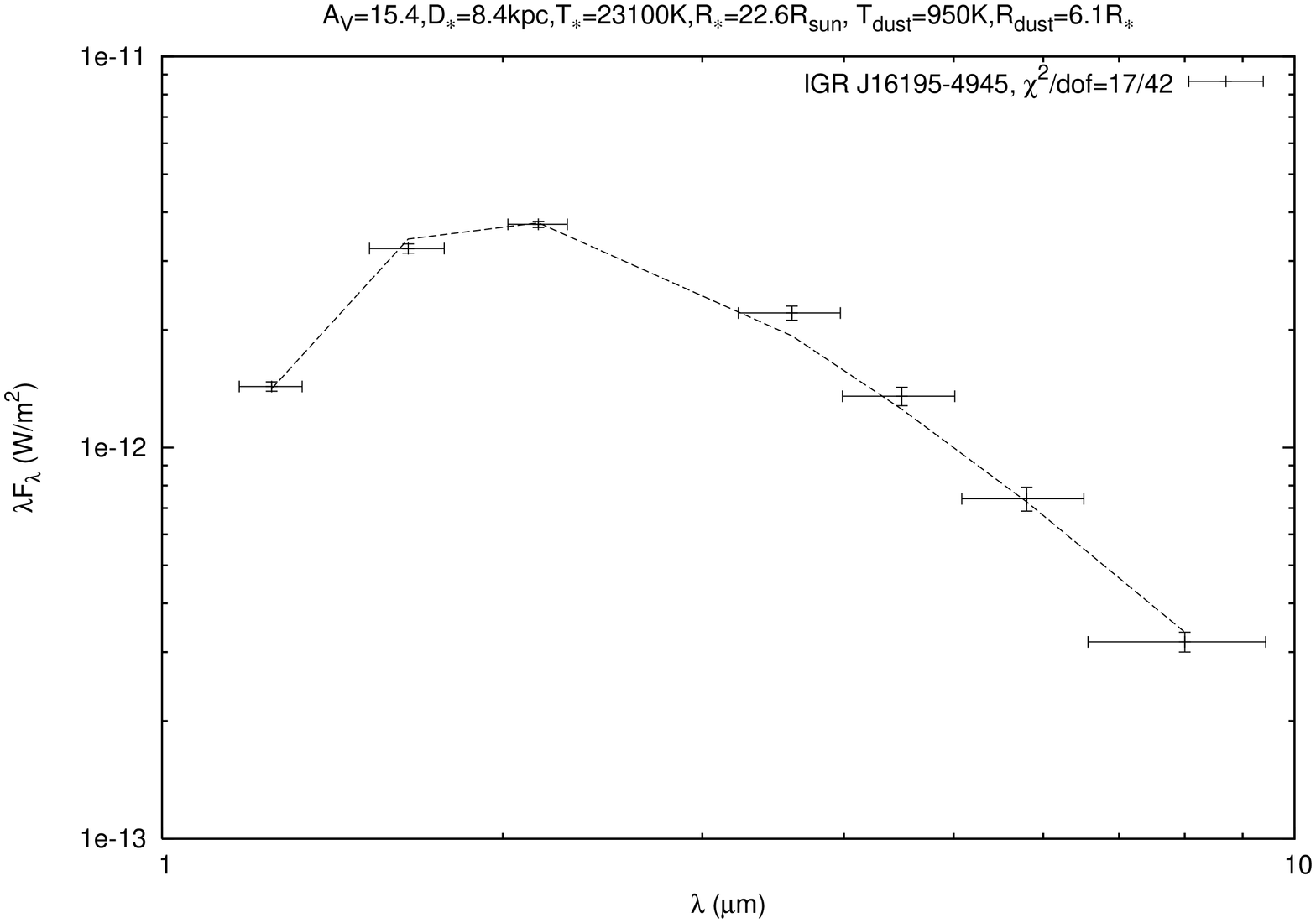,angle=-90.,width=10.cm}}
\caption[]{\label{figure:igrj16195} NIR to MIR SED of IGR J16195-4945,
including data from ESO/NTT and {\it Spitzer} (GLIMPSE survey). 
The observations fitted with both a model of a companion star 
(taking usual parameters of an O/B star) and simple dust component
allowed us to derive the following parameters:
i) for the companion star: temperature
T~$=23100$~K and radius R$_{star} = 22.6 R_{\odot} = 15
\times 10^6$~km;
ii) for the dust component: temperature T $=950K$, radius
R~$= 6.1 R_{star} = 95 \times 10^6$~km.
The absorption we derived was A$_v = 15.4 \mags$, and the distance
 D~$=8.4$~kpc. The $\chi^2$/dof value of the fit is 17/42.
The main result given by this SED is that 
in the case of IGR J16195-4945, as for IGR J16318-4848, there is a need for
an extra (e.g. dust) component in order to fit its SED. }
\end{figure}

We already showed the fit of IGR J16318-4848 in Figure \ref{figure:igrj16318}.
We show the fit of IGR J16195-4945 SED from NIR to MIR
wavelengths, including data from ESO/NTT and {\it Spitzer} (GLIMPSE survey) 
in Figure \ref{figure:igrj16195}.  We fitted the
observations with a model of a companion star (taking usual parameters
of an O/B star) and simple spherical dust component.  We found a temperature of
T~$=23100$~K, radius R$_{star} = 22.6 R_{\odot} = 15
\times 10^6$~km for the parameters of the companion star,
and a dust component with the following parameters: T~$=950$~K, radius 
R~$= 6.1 R_{star} = 95 \times 10^6$~km and A$_v = 15.4 \mags$.  The
derived distance was of D~$=8.4$ kpc.  The $\chi^2$/dof of the fit was
of 17/42, which shows the good quality of the fit.  What is important
in this result is that in the case of IGR J16195-4945, as for IGR
J16318-4848, there is the need for an extra (e.g. dust)
component. Again, as for IGR J16318-4848, the 
extension of this dust component seems to suggest that
it is enshrouding the whole binary system, perhaps as would do a
cocoon of dust.

We have to point out that we are
 more cautious in our conclusions about IGR J16195-4945
than with IGR J16318-4848, since i) the former was not detected with
VISIR, but only with SPITZER, and ii) 
the ESO/NTT optical magnitudes seem to be those of a blended object
(see \citeauthor{tomsick:2006}    \citeyear{tomsick:2006} 
and \citeauthor{tovmassian:2006} \citeyear{tovmassian:2006}).
However, both sources seem to be very similar, since they exhibit the
same temperature ($\sim 23000$ K) and are highly obscured sources:
they both exhibit absorption of A$_V \sim 17 \mags$ in the optical,
and their column density derived from X-ray observations is
respectively $\nh = 2.1 \times 10^{24} \cmmoinsdeux$ for IGR
J16318-4848 and $\nh \sim 10^{23} \cmmoinsdeux$ for IGR J16195-4945.
In fact, the case of IGR J16195-4945 is extremely interesting, since
it would look very much like IGR J16318-4848 if it were located at the
same distance (our fits suggest that IGR J16195-4945 is 7 times more
distant than IGR J16318-4848).  Therefore, the parameters derived from
our fits suggest that IGR J16195-4945 should not be visible in
optical, and this result is consistent with the conclusion by
\cite{tovmassian:2006} that the optical source observed in
\cite{tomsick:2006} is not the {\it INTEGRAL} source but a foreground
object.

Another important point is that the fits are very dependent 
on the absorption correction used, and the sources which exhibit
a high absorption seem to be better fitted with the absorption
correction given in \citet{indebetouw:2005}
than the one of \citet{rieke:1985}, probably because the former has
been calibrated using
 more recent MIR data taken from {\it Spitzer} observations.
In summary, if the observations and the fits leave no place to any doubt 
about the presence of dust around IGR J16318-4848, the situation is less
clear for IGR J16195-4945, even if the parameters derived by fitting
 the observations suggest its presence.

Therefore, in two cases only, concerning the sources IGR J16318-4848
and IGR J16195-4945, the presence of cold dust is required by the fits.
In this context, IGR J16318-4848 proves once again that it is an
extraordinary source among other {\it INTEGRAL} sources, and that there is
much more absorbing material around this source than around the
others.  Therefore IGR J16318-4848, and probably also IGR J16195-4945,
remain exceptional cases, which might probably deserve to constitute a
class by themselves!

\begin{table*}
  \begin{center}
    \caption{Summary of MIR observations of newly discovered {\it INTEGRAL} sources.
We give in this Table the name of the sources, their coordinates,
their type (SFXT or OBS --obscured source--), their spectral type (SpT),
the reference (Ref) about the spectral type, and their MIR
magnitudes in the PAH1 ($8.59 \microns$), PAH2 ($11.25 \microns$)
and Q2 ($18.72 \microns$) filters.
The classification as SFXT is still subjective, since we miss some
accurate observations on a long-term scale for most of the sources. 
The spectral types come from optical/NIR spectroscopy,
reported in the following references: 
c: Chaty et al. in prep.,
f: \cite{filliatre:2004},
i: \cite{intzand:2006},
n1: \cite{negueruela:2005},
n2: \cite{negueruela:2006a},
p: \cite{pellizza:2006},
t: \cite{tomsick:2006},
z: \cite{zurita-heras:2006}.
}\vspace{1em}
    \renewcommand{\arraystretch}{1.2}
    \begin{tabular}{ccccccccc}
\hline
Sources&$\alpha$ (J2000)&$\delta$ (J2000)&Type&SpT&Ref&PAH1&PAH2&Q2\\
\hline
IGR J16195-4945&16 19 32.20&-49 44 30.7&OBS&OB&t&$<$ 6.12&$<$ 7.83&$<$ 50.25\\
\hline
IGR J16207-5129&16 20 46.26&-51 30 06.0&&&t&22$\pm$1.4&9.4$\pm$1.0&$<$ 53.37\\
\hline
IGR J16318-4848&16 31 48.60&-48 49 00.0&OBS&sgB[e]&f&409$\pm$2.4&322$\pm$3.26&172$\pm$14.9\\
\hline
IGR J16320-4751&16 32 01.90&-47 52 27.0&&OB&c&12.1$\pm$2.67&6.3$\pm$1.84&\\
\hline
IGR J16358-4527&16 35 53.80&-47 25 41.1&&&&$<$ 6.84&&\\
\hline
IGR J16418-4532&16 41 51.00&-45 32 25.0&SFXT&&&$<$ 5.83&&\\
\hline
IGR J16465-4507&16 46 35.50&-45 07 04.0&SFXT&B0.5I&n1&8.69$\pm$1.77&4.7$\pm$0.9&\\
\hline
IGR J16479-4514&16 48 06.60&-45 12 08.0&SFXT&OB&c&12$\pm$1.3&7$\pm$1.7&\\
\hline
IGR J17252-3616&17 25 11.40&-36 16 58.6&&OB&z&6$\pm$0.6&2$\pm$0.4&\\
\hline
IGR J17391-3021&17 39 11.58&-30 20 37.6&SFXT&O8Iab(f)&n2&70.2$\pm$1.61&46.5$\pm$2.64&\\
\hline
IGR J17544-2619&17 54 25.28&-26 19 52.6&SFXT&O9Ib&p&36$\pm$2.77&20.2$\pm$2.07&\\
\hline
IGR J17597-2201&17 59 45.70&-22 01 39.0&&&&$<$ 6.12&&\\
\hline
IGR J18027-2016&18 02 42.00&-20 17 18.0&&sgOB&c&$<$ 6.00&&\\
\hline
IGR J19140+0951&19 14 04.23&+09 52 58.3&&OB?&c&35$\pm$1.4&19$\pm$1.4&\\
\hline
\end{tabular}
\label{tab:mirobs}
  \end{center}
\end{table*}

\section{Discussion and conclusions} \label{discussion}

Now, the question which remains is: ''what are these sources?''.  80\%
of these newly discovered {\it INTEGRAL} sources are HMXBs, hosting
compact objects (probably neutron stars) 
orbiting around O/B supergiant secondaries.  These
systems are wind accretors, and exhibit a more or less substantial
extra absorption.  Obscured sources and SFXTs share similar
properties, however, they are not the same type of sources, mainly
because this excess in absorption does not seem to have the same
origin in both classes of sources.  For instance, the excess of
absorption is caused by two different phenomena in the case of the
highly obscured sources, such as IGR J16318-4848.  Indeed, in this
case, the observations from high energy to MIR domains suggest that
there is some absorbing material concentrated around the compact
object, and also some dust, cold gas, or even a cocoon of dust,
enshrouding the whole binary system.  On the contrary, in the case of
SFXTs, such as IGR J17544-2619, the presence of the absorbing
material seems
concentrated around the compact object only, and MIR observations show
that there is no need of any other absorbing material around the whole
system.

We can therefore try to distinguish the nature of both classes by 
speculating on the geometry of these systems:

\begin{itemize}

\item The highly obscured sources (for which the archetype is
IGR J16318-4848) are characterised by the presence of absorbing
material both around the compact object and around the whole binary
system.  Their characteristics might be explained by the presence of a
compact object (neutron star or black hole) 
orbiting within the dense wind surrounding the companion star.

\item The SFXTs (for which the archetype is IGR J17544-2619) are
characterised by fast X-ray outbursts, and the presence of a
supergiant companion star.  Their characteristics might be explained
by the presence of a compact object (neutron star or black hole) 
located on a wide orbit around the
companion star, and it is probably when the compact object penetrates
the envelope of the star that outbursts are caused.

\end{itemize}

Therefore the X-ray transient or persistent nature of these
sources might be related to the geometry of these systems.
Obviously the confirmation of this view will probably be given by the
knowledge of their orbital periods.  Many questions are still open, and
most of them are related to the presence of the MIR excess in these sources.
For instance, no radio emission has been detected in any of these
systems, while it is commonly detected among high energy binary
systems, therefore it seems that there is something special here again
with these sources. One possibility is that the dust might prevent the
triggering of the jets. But in order to answer to this question,
we will need to better characterise the dust, its temperature, 
composition, geometry, extension around the system, etc...
And also, we need to investigate where this dust or cold gas
comes from...
But probably the most important question is: is this unusual
circumstellar environment due to stellar evolution OR to the binary
system itself?  We are now facing a dominant population of high energy
binary systems born with two very massive components. These systems
are probably the primary progenitors of NS/NS or NS/BH mergers.  
There is therefore the possibility that they are related with short/hard
gamma-ray bursts, and also that they might be good candidates of
gravitational wave emitters.

To summarise, a new population of sources has been recently
revealed by {\it INTEGRAL}, and it appears that a careful study of this new
population might provide hints on the geometry of high energy 
binary systems, and a better understanding of the evolution of these
systems.
Our final word will be that, because they are obscured, 
the ''Norma arm'' sources can only be studied in the 
high-energy and infrared domains.
A joint study with multiwavelength high-energy, optical, NIR, MIR
(and radio) observations is therefore necessary, especially during bursts
of these sources.

\section*{Acknowledgements}

Based on observations collected at the European Southern Observatory,
Chile (observing proposals ESO N$\adeg$ 70.D-0340, ESO 71.D-0073, 075.D-0773
and 077.D-0721).  SC is grateful to Leonardo Pellizza and
J\'er\^ome Rodriguez for a careful
rereading of the manuscript.  SC would like also to thank the
organisers for the opportunity to report on these exciting results on
newly discovered {\it INTEGRAL} sources, and also for a very nice
organisation of this workshop, fruitful to arise scientific
discussions and new ideas. \\
{\cyr Bolshoe spasibo, Poka!}


\bibliographystyle{/Users/chaty/Library/Texmf/Bibtex/aa}



\end{document}